# Fast and accurate Deflectometry with Crossed Fringes

Yuankun Liu[a,b]*, Evelyn Olesch[b], Zheng Yang[b], and Gerd Häusler[b]

(*a*Opto-Electronic Department, Sichuan University, Chengdu 610065, China)
(*b*Institute of Optics, Information and Photonics, University of Erlangen-Nuremberg, Erlangen, 91058, Germany)

*Corresponding Author: yuankun_liu@hotmail.com*

Phase Measuring Deflectometry (PMD) acquires the two components of the local surface gradient via a sequence of two orthogonal sinusoidal fringe patterns that have to be displayed and captured separately. We will demonstrate that the sequential process (different fringe directions, phase shifting) can be completely avoided by using a cross fringe pattern. With an optimized Fourier evaluation, high quality data of smooth optical surfaces can be acquired within one single shot.

The cross fringe pattern allows for one more improvement of PMD: we will demonstrate a novel phase-shift technique, where a *one-dimensional* N-phase shift allows for the acquisition of the *two orthogonal phases*, with only N exposures instead of 2N exposures. So, PMD can be implemented by a *one-dimensional* translation of the fringe pattern, instead of the common two-dimensional translation, which is quite useful for certain applications.

Absolute Phase Measuring Deflectometry (PMD) [1-3], as similar methods [4-7], is based on the observation of mirror images of remote patterns, using the object under test as a mirror. The mirrored patterns are distorted, depending on the shape of the object. Using sinusoidal fringes and a thoroughly calibrated system, the local slopes can be acquired by standard phase shifting techniques. An integration algorithm [8] will be used to reconstruct the shape of the object.

In order to obtain the two components of the local surface gradients, two orthogonal sinusoidal fringe patterns are displayed separately, each with a sequential phase shift. Eight exposures are necessary for the (common) four-phase-shift measurement. Can we avoid the multiple exposures and achieve "single-shot" deflectometry? This would be quite advantageous, specifically for measurements during the manufacturing process [9]. We could measure moving objects or just measure faster.

However, this is impossible: generally, one single exposure does not deliver sufficient information about the unknowns - reflectivity, ambient light, local phase [10]. Takeda [11] suggested a workaround for fringe projection triangulation, based on single side band filtering. However, this method is only applicable if the image bandwidth is less than 1/3 of the camera bandwidth (the factor 3 is buying us the three unknowns). In practice, the bandwidth limit is a serious drawback for the majority of 3D objects.

However, there is one important application with extremely low bandwidth: deflectometry at smooth optical surfaces, such as eye glasses. We will demonstrate that for this application we can overcome the sequential acquisition of the two gradient components by crossed fringes, and the sequential phase shift by a properly adapted single sideband Fourier evaluation. The optimized Fourier evaluation largely reduces artifacts at the object rim and allows for comparably high lateral resolution. In principle, the measurement can be as fast as one single camera exposure. Even more important is that we can measure dynamical processes, such as moving objects or choppy liquid surfaces.

Cross fringe illumination offers one more option: we will demonstrate that it is possible to implement phase shifting for both (horizontal and vertical) fringe components by a lateral shift of the pattern in *only one dimension.* For an N-phase-shift only N exposures are necessary, instead of the common 2N exposures. Nevertheless the method is highly accurate, and is faster than other phase-shift techniques..

Metrology by cross fringe patterns has been suggested earlier: Huang [12] proposed dynamic (single shot) 3D sensing of a liquid surface. The fringe phase is acquired by a Fourier method similar to Takeda's method. The difference to our method results from the application: The liquid displays a low height range (~20μm) and there is no visible object boundary, so Huang's implementation does not have to take account for lateral resolution, accuracy and filtering artifacts.

We will now explain our single-shot method: The image intensity of the additive cross fringe pattern is described by

$$I(x,y) = a(x,y) + b_1(x,y)\cos[\phi_x(x,y) + \delta_x] \\ + b_2(x,y)\cos[\phi_y(x,y) + \delta_y] \quad (1)$$

where (*x,y*) are the two-orthogonal directions of the screen, *a(x,y)* is the bias, $b_1(x,y)$, $b_2(x,y)$ are the modulations (~local reflectivity). $\phi_x$ and $\phi_y$ are the two orthogonal phase distributions. Equation (1) asks for five unknowns (instead of three, as for triangulation), so at least five exposures have to be taken. This could principally be done by introducing sufficient phase shifts $\delta_x$ and $\delta_y$, which will be explained later.

As mentioned above, we have a fortunate situation: eye glasses (and most optical surfaces) are smooth and we can extract $\phi_x$ and $\phi_y$ (~the local slope components) from only one exposure, via Fourier filtering. Figure 1 displays the image of the fringes, deformed by reflection at the eye glass. A marker (in the circle), is designed to identify the fringe order. Then Gerchberg iteration [13] is used to overcome the local accuracy reduction introduced by the marker. Figure 2 shows the Fourier spectrum (magnitude, logarithmic scale) of the camera image and as well the rims of the Hanning filter windows.

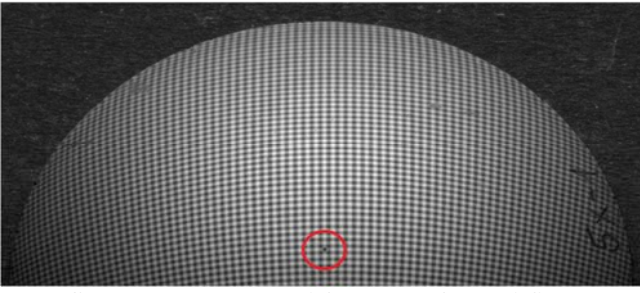

Fig.1. Camera image captured from cross fringe deflectometry. The object is an eye glass surface with ~8.5 D. The marker helps to identify the fringe number.

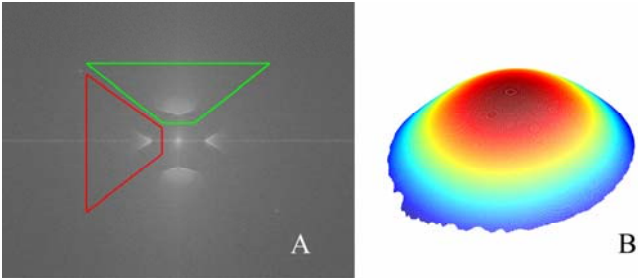

Fig.2. a) Fourier spectrum (log scale) with the rims of the Hanning filter windows; b) reconstructed surface.

How to choose the optimal grid frequency $\nu_G$? Since the bandwidth of the camera image is very small (see as well Fig. 2a), $\nu_G$ can – and should – be as large as possible.

Then the carrier frequency will be far away from the base term and far away from the $1/\nu$ term generated by the (not band limited) edges of the object. So the band-pass filter can be wide as shown in Fig. 2a, to allow for a transmission of the required phase information with high lateral resolution. At the same time, the wide filter will produce less visible artefacts at the object edges. We found an optimal grid frequency $\nu_G = (8 \text{ pixel})^{-1}$. Higher frequencies are possible, but the results display more noise, due to less contrast in the camera image.

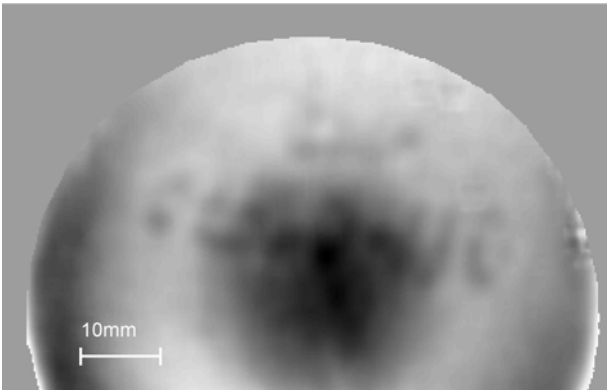

Fig.3. Curvature, calculated from the measured slope. The variation of the curvature range is from 9.2 D (bright) to 8.2 D (dark). The marks display an about 10nm local material wear by an earlier (cleaned) felt pen marking.

The considerations above are confirmed by Fig. 3: Figure 3 displays an intensity encoded curvature map of the measured object. The observable letter structure was indeed caused by letters written on the glass, with a felt pen. Although the letters were wiped away after the writing, a local material wear in the range of 10 nm remained, its curvature variation can be seen in Fig. 3.

The experiments display that a measurement of real optical surfaces is possible, in real time, with relatively high lateral resolution, low noise and low filtering artefacts.

On the other hand, real objects are not truly band limited, so measuring errors at edges or scratches are principally inevitable due to the Fourier filtering. If real-time measurement is not required, it turns out that the problems mentioned above can be avoided by a phase-shift technique. We will demonstrate that with cross fringe illumination this is even easier and faster than with standard deflectometry.

Canabal [14] proposed a cross fringe pattern for Moiré deflectometry, with a phase-shift in *two* directions separately, so some pixels are always modulated with low intensity for one direction. This could be avoided by introducing an extra N-shift for each direction with a π shift in the other direction. The authors used this technique as well with a multiplicative cross fringe pattern for Moiré deflectometry [15]. They need 2*2*N exposures and a two-dimensional grid translation.

We will demonstrate a novel phase-shifting strategy, which requires only N exposures and an only one-dimensional grid translation. Our idea was inspired by Liu [16], who combined two frequencies in one grid pattern, for fringe projection triangulation. In order to decipher the corresponding phases, Liu used different phase steps for each frequency.

The five unknowns in Eq. (1) require at least five phase-shifts. It turns out that the two orthogonal phases ($\phi_x$ and $\phi_y$) can be calculated from Eq. (1) *only* if the phase steps for the x and y components are different.

We re-write Eq. (1):

$$I_n = a + b_1 \cos[\phi_x(x,y) + n\frac{2\pi}{N_1}] + b_2 \cos[\phi_y(x,y) + n\frac{2\pi}{N_2}]$$

$$N_1 = 5,6,7\cdots$$

$$N_2 = \frac{N_1}{k}, k = 2,3\cdots, N_1 - 2$$

(2)

where $2\pi/N_1$ is the phase step for the x direction, $N_1$ is the number of total exposures, and $N_2$ depends on $N_1$ and $k$. $N_1$ samples are acquired in $k$ periods for the y direction, i.e. the shift angle is $2k\pi/N_1$ and the total phase shift angle is $2k\pi$. For example, $N_1$=5 and $k$=2 mean five samples in two periods. The two orthogonal phases can be extracted by

$$\phi_x(x,y) = \tan^{-1} \frac{\sum_{n=0}^{N_1-1} I_n \sin(n\frac{2\pi}{N_1})}{\sum_{n=0}^{N_1-1} I_n \cos(n\frac{2\pi}{N_1})} \qquad (3)$$

$$\phi_y(x,y) = \tan^{-1} \frac{\sum_{n=0}^{N_1-1} I_n \sin(n\frac{2\pi}{N_2})}{\sum_{n=0}^{N_1-1} I_n \cos(n\frac{2\pi}{N_2})}$$

Note that $N_2$ should be not less than three when it is an integer. Figure 4 illustrates the method for an example $N_1$=5 and $k$=2.

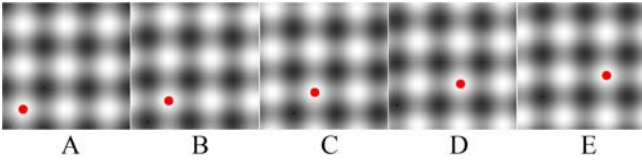

Fig. 4. The virtual movement of the cross fringe pattern for a five-phase-shift.

In order to clarify the (virtual) pattern translation, a marker is drawn at a fixed phase. Figure 4 shows that the pattern is shifted by two periods horizontally and by one period vertically (depending on k and $N_1$), but only in one dimension.

Compared to the sequential two-dimensional phase-shifting [14], there are no fixed low intensity pixels in our method, and we only need $N_1$ exposures, instead of $2*2N_1$. We compared our method with the standard PMD, see Fig. 5 and Table 1. The measurements are performed with a grid period of 8 pixels. The additive cross fringe pattern combines two orthogonal fringes, and the modulation for each component is only 50% of the standard PMD. Therefore, we expect a larger repeatability error for the cross fringe measurement, as shown in Fig. 5.

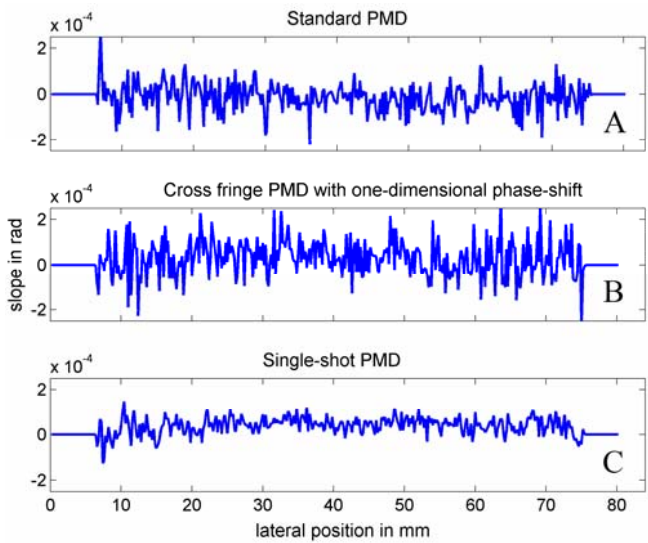

Fig.5. Repeatability errors, a) standard PMD, b) cross fringe PMD with one-dimensional phase shift, c) single-shot PMD.

For cross fringe PMD, we also tested the sequential phase shift in two directions (2*3 and 2*5 exposures), see Table 1. It is evident that, with the same number of exposures 2*N, higher modulation will deliver higher precision, and our one-dimensional phase shift is better than the sequential two-dimensional phase shift. In any case, the accuracy can be improved by increasing the number of phase steps as shown in Table 1.

Table 1. Slope repeatability (in arcsec) of different methods with different number of exposures

| Exposures | 1 | 6 | 10 |
|---|---|---|---|
| Single-shot PMD | 7.8 | - | - |
| Standard PMD | - | 11.9 | 9.4 |
| Cross fringe PMD with one-dimensional phase-shift | - | 15.8 | 11.1 |
| Cross fringe PMD with two-dimensional phase-shift | - | 24.5 | 18.7 |

It is eye catching that the single-shot PMD displays considerably less noise than all other methods. Of course, this is due to the band pass filter and we have to pay for the low noise by reduced lateral resolution.

We summarize: when the tested object is smooth (as most optical surfaces are), we can implement a single-shot PMD, with high precision, relatively high lateral resolution, and the option to measure dynamic processes.

The Fourier Fringe analysis inevitably causes residual errors at boundaries and fine details. If we do not need the single-shot advantage, a proper cross fringe projection allows for an efficient phase shifting strategy. With a one-dimensional phase-shifting we avoid the residual errors of the single shot method and achieve high accuracy.

Note that we precisely rectify the nonlinearity, before the measurement, because otherwise the two additive frequency components will introduce mixed frequencies, which will increase the difficulties for the filtering in single-shot applications and which will cause residual errors in the phase-shifting technique.

Eventually, we want to mention a great practical advantage of cross fringe deflectometry with one-dimensional phase shifting:

There are applications where there is no electronic imager available e.g., for very big objects [17] and for deflectometry with ultraviolet light [18]). The UV deflectometry is used to avoid parasitic backside reflections. In these applications, the implementation via an only *one-dimensional* translation of a physical grid mask is extremely useful.